\documentclass{elsart}

\usepackage{graphicx}

\usepackage{amssymb,amsmath}

\begin{document}

\begin{frontmatter}



\title{Recent progress in first-principles studies of magnetoelectric
multiferroics}


\author{Claude Ederer}\ead{ederer@mrl.ucsb.edu}, \author{Nicola
A. Spaldin}\ead{nicola@mrl.ucsb.edu}

\address{Materials Department, University of California, Santa
  Barbara, CA 93106}

\begin{abstract}
Materials that combine magnetic and ferroelectric properties have
generated increasing interest over the last few years, due to both
their diverse properties and their potential utility in new types of
magnetoelectric device applications. In this review we discuss recent
progress in the study of such magnetoelectric multiferroics which has
been achieved using computational first-principles methods based on
density functional theory. In particular, we show how first-principles
methods have been successfully used to explain various properties of
multiferroic materials and to predict novel effects and new systems
that exhibit multiferroic properties.
\end{abstract}

\end{frontmatter}

\section{Introduction}
\label{intro}


Magnetoelectric multiferroics are materials which exhibit both
magnetic order and ferroelectricity in the same phase. Such materials,
although rare, have been known since the 1960s and various early
review articles systematically classified their properties and
behavior
\cite{Skinner:1970,Schmid:1973,Smolenskii/Chupis:1982,Schmid:1994}. In
the last few years the field of multiferroic materials has seen a
tremendous boom, initiated in part by computational first-principles
studies explaining the basic physics underlying their scarcity
\cite{Hill:2000}. The subsequent flurry of research activity led to
the discovery of several new multiferroic materials, as well as the
identification of various strong coupling effects between their
magnetic and ferroelectric degrees of freedom
\cite{Fiebig_et_al:2002,Wang_et_al:2003,Kimura_et_al_Nature:2003,Kimura_et_al_PRB:2003,Hur_et_al:2004,Lottermoser_et_al:2004}.
The prospect of using these coupling effects for new types of devices
in microelectronics and digital memory applications has in turn
spawned further interest.

This newer work on magnetoelectric multiferroics was recently reviewed
in Ref.~\cite{Fiebig:2005}, with an emphasis on experimental and
technological studies. The purpose of the present article is to
provide an update on the recent activities in first-principles studies
of magnetoelectric multiferroics, with the goal of illustrating the
utility of these methods both in explaining unusual behavior in
existing multiferroics, and in predicting new materials and novel
effects. An extensive review of density functional theory-based
studies of multiferroic materials before 2002 was given in
Ref.~\cite{Hill:2002}. Since then, first-principles studies of
multiferroic materials have made considerable progress; here we
summarize these recent developments and point out possible directions
for future work.


Ref.~\cite{Hill:2002} reviewed two major questions which were
successfully answered by first-principles electronic structure
calculations. The first question, ``Why are there so few
magnetoelectric multiferroics?''  \cite{Hill:2000} is now understood
to result from a chemical incompatibility between magnetism and
conventional ferroelectricity. This incompatibility is related to the
fact that the most common mechanism for ferroelectricity in
perovskite-structure oxides involves the presence of a transition
metal cation on the perovskite $B$ site with a formal $d^0$ electron
configuration.  In fact, first-principles electronic structure
calculations have played a major role in elucidating the role of the
$d^0$ electron configuration in the ferroelectric instability in
perovskite oxides \cite{Cohen:1992,Cohen:2000,Filippetti/Hill:2002}.
On the other hand, for magnetism to occur in such transition metal
oxides a partially filled $d$ shell is indispensable. The existence of
at least some known materials where magnetism and ferroelectricity
coexist then poses the second question addressed in
Ref. \cite{Hill:2002}: ``Why are there {\it any} magnetoelectric
multiferroics?''  \cite{Hill/Filippetti:2002}. Clearly, if the
ferroelectricity in a multiferroic is caused by a $d^0$ cation, a
different cation is needed to introduce the magnetism; alternatively,
if only magnetic cations with partially filled $d$ shells are present,
a new mechanism for ferroelectricity is required. To our knowledge,
all known multiferroics adopt the latter scenario. An alternative
mechanism for ferroelectric ``off-centering'', which was discussed in
Ref.~\cite{Hill:2002}, is the stereochemically active lone pair in
cations such as Pb$^{2+}$ or Bi$^{3+}$. Again, first-principles
calculations were invaluable in showing that this mechanism is
responsible for the ferroelectricity in multiferroics such as
BiMnO$_3$ or BiFeO$_3$ \cite{Seshadri/Hill:2001,Neaton_et_al:2005}. In
these systems, the Bi$^{3+}$ cation has a formal valence electron
configuration of 6$s^2$6$p^0$ and the energy of the system can be
lowered by off-centering of the Bi ion with respect to its oxygen
surrounding, which leads to a hybridization of both Bi 6$s$ and 6$p$
states with O 2$p$ orbitals and a localization of the lone pair on one
side of the Bi ion \cite{Seshadri/Hill:2001}.


The organization of this article is as follows. In the following
section we give a brief overview of some methodological aspects of
first-principles electronic structure calculations that are especially
relevant to the study of magnetoelectric multiferroics. Some recent
methodological developments have contributed to the current boom in
magnetoelectric multiferroics, since they facilitate a realistic
description of the electronic structure of multiferroic materials and
allow the calculation of quantities that are difficult to access
experimentally.

Following this brief methodological part, we discuss how
first-principles calculation have led to substantial progress in the
understanding of multiferroic materials during the last few years. We
begin with cases in which first-principles calculations have explained
properties of existing multiferroics. As examples we discuss the
identification of the mechanism responsible for ferroelectricity in
the hexagonal manganite YMnO$_3$, a material which is representative
for a whole class of multiferroics, as well as the resolution of
controversy surrounding confusing and contradictory experimental data
in multiferroic BiFeO$_3$. Following that, we show how
first-principles calculations have made important predictions in
advance of their experimental observation.  Here, we also discuss two
examples: The identification of a mechanism for coupling between
magnetic and ferroelectric order parameters that could facilitate the
intriguing effect of electric-field induced magnetization switching,
and the {\it ab initio} design of new multiferroic materials with
superior materials properties. Finally, we give an outlook on future
research in the field, and point out where first-principles methods
can further contribute to the study of magnetoelectric multiferroics.

\section{Computational tools}
\label{sec:method}

In this section we briefly summarize recent developments in electronic
structure methods which are particularly important for, and in one
case were motivated by, the study of multiferroic systems. For a
complete description of density-functional-theory-based
first-principles methods we direct the reader to one of the excellent
texts which are available,
e.g. Refs.~\cite{Jones/Gunnarsson:1989,Martin:Book}.

The first electronic structure calculations of multiferroic systems were
performed using the local spin density approximation (LSDA) to density
functional theory \cite{Jones/Gunnarsson:1989}. Although the LSDA is a
well-established technique that continues to make invaluable contributions to
our understanding of ferroelectrics (as well as many other materials), it has
a number of limitations, among them its well-known underestimation of the size
of the band gap for most insulating materials. For systems containing
localized electrons with strong Coulomb correlations such as transition metal
oxides the use of the LSDA can be totally inadequate, since it can lead to
metallic solutions for systems that are known experimentally to be insulators
\cite{Terakura_et_al:1984}. In the case of magnetic ferroelectrics this is
particularly problematic, since it prohibits the calculation of the
spontaneous ferroelectric polarization and other indicators for structural
instabilities, such as Born effective charges
\cite{Ghosez/Michenaud/Gonze:1998}. Therefore, the most important recent
progress in first-principles methodology, with respect to the study of
multiferroic materials, has been the development of methods that facilitate a
realistic description of the electronic structure of magnetic systems
containing strongly localized $d$ or $f$ electrons.

Two methods that attempt to cure some of the deficiencies of the LSDA in the
treatment of localized electrons were developed recently and have been applied
to study multiferroic materials: the so-called ``LSDA plus Hubbard $U$''
method (LSDA+$U$) \cite{Anisimov/Aryatesiawan/Liechtenstein:1997} and a
self-interaction corrected pseudopotential method (``pseudo-SIC'')
\cite{Filippetti/Spaldin:2003}. Although the physical motivation behind these
two methods is different, they lead to very similar results in practice.

The development of the pseudo-SIC method was actually motivated by the
inability of the LSDA to produce a band gap for multiferroic YMnO$_3$
(see section~\ref{YMnO3} below). In general, self-interaction
corrected (SIC) methods are based on the well-known fact that the
failures of the LSDA can be, at least in part, attributed to the
presence of the self-interaction (SI) in the LSDA energy functional,
that is the interaction of an electron charge with the Coulomb and
exchange-correlation potential generated by the same electron
\cite{Jones/Gunnarsson:1989}. The SI is strong for spatially localized
electrons such as $3d$ and $4f$ electrons, and leads to an
underestimation of the binding energies, on-site Coulomb energies and
exchange splittings of the corresponding states, while the
hybridizations of cation $d$ and anion $p$ states and the
corresponding band widths are usually overestimated. Possible
strategies for eliminating the SI in density functional theory are
long-standing issues, but often require a large computing effort even
for materials with small unit cells. In the pseudo-SIC method
\cite{Filippetti/Spaldin:2003}, the SI correction is calculated for
the corresponding atoms (as in Ref. \cite{Perdew/Zunger:1981}), and is
then scaled by the electron occupation numbers of specific atomic-like
orbitals calculated self-consistently within the crystal
environment. This approach can be efficiently implemented within a
standard pseudopotential method and allows the SI originating from
localized, hybridized, or completely itinerant electrons to be
discriminated, which permits the treatment of insulating as well as
metallic compounds, with minimal computational overhead beyond the
LSDA \cite{Filippetti/Spaldin:2003}.

The motivation behind the development of the LSDA+$U$ method is somewhat
different, although in practice this method leads to very similar results as
the pseudo-SIC method. In the LSDA+$U$ method the Coulomb interaction between
specific localized orbitals is explicitly taken into account by adding a term
to the total energy that resembles the well-known``Hubbard Hamiltonian'' used
to describe Mott-Hubbard metal-insulator transitions
\cite{Anisimov/Zaanen/Anderson:1991}. The resulting potential leads to an
energy splitting between occupied and unoccupied states. The Hubbard
parameter, $U$, which describes the strength of the screened electron-electron
interaction, can in principle be calculated {\it ab initio} but in practice it
is often treated as an empirical parameter. Simultaneously, a
``double-counting correction'' is subtracted from the total energy in order to
account for those interaction effects between the localized electrons that are
already included in the conventional LSDA. Different forms for this correction
term are in use, some of which are in part also designed to make the resulting
energy functional exhibit certain desirable properties
\cite{Solovyev/Dederichs/Anisimov:1994}. These ambiguities and the partly
empirical construction of the Hubbard correction is somewhat unsatisfying and
the resulting method is strictly speaking not purely first
principles. Nevertheless, the LSDA+$U$ method is nowadays used more and more
routinely in electronic structure calculations for $d$ and $f$ electron
systems and has undoubtedly led to considerable progress in the field.

Simultaneously with the tremendous improvements in the first-principles
description of magnetic insulators provided by the LSDA+U and pseudo-SIC
methods, the so-called ``modern theory of polarization'' has revolutionized
the first-principles study of ferroelectric materials
\cite{King-Smith/Vanderbilt:1993,Vanderbilt/King-Smith:1993,Resta:1994}. This
theory provides both conceptual understanding of the spontaneous polarization
in bulk crystals, and computational tools to calculate it from the Kohn-Sham
wave-functions. Complications in calculating the electric polarization in bulk
crystals arise from the periodic boundary conditions employed in most solid
state calculations. This periodicity leads to an ambiguity in the value of the
polarization modulo an integer multiple of the so called ``polarization
quantum'', $\frac{eR}{V}$, such that
$$
P = P' + n \frac{eR}{V} \quad ,
$$ 
where $P$ and $P'$ are two allowed values of the polarization, $n$ is an
integer number, $e$ is the electronic charge, $V$ is the unit cell volume, and
$R$ is a primitive lattice vector of the underlying Bravais lattice. The
polarization of an infinite periodic crystal is therefore represented by a
lattice of values, the so-called ``polarization lattice''. In an experiment
only polarization changes can be measured and the corresponding value has to
be calculated as a difference between two points of the polarization lattices
corresponding to the initial and final states of the system. The corresponding
change in polarization is uniquely defined as long as the system stays
insulating for all intermediate states transforming the initial to the final
state.

\begin{figure}
\centerline{\includegraphics[width=0.7\textwidth]{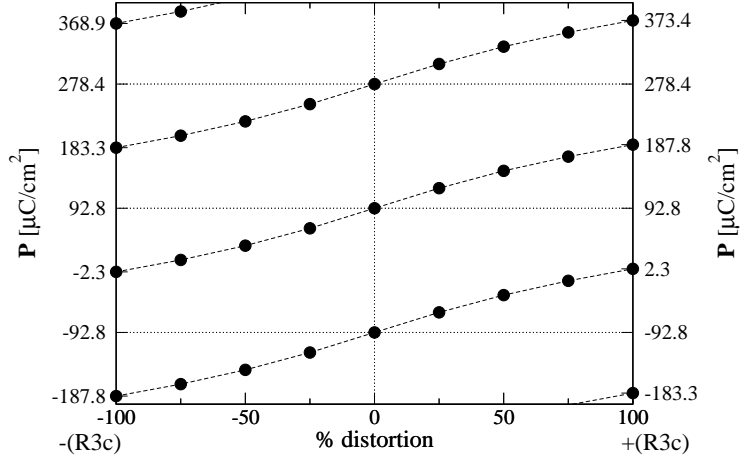}}
\caption{ Change in polarization $P$ along a ferroelectric switching path for
  multiferroic BiFeO$_3$. The possible values of $P$ for fixed distortion
  differ by multiples of the polarization quantum, here
  $185.6~\mu$C/cm$^2$. From Ref. \cite{Neaton_et_al:2005}. Copyright (2005) by
  the American Physical Society.}
\label{fig:path_endpt}
\end{figure}

Figure~\ref{fig:path_endpt} shows the evolution of the polarization
lattice in multiferroic BiFeO$_3$ calculated for different points
along a ferroelectric switching path leading from the negatively
polarized ferroelectric state ($-$100\% distortion), through the
corresponding centrosymmetric structure (0\% distortion), to the
positively polarized state (+100\% distortion)
\cite{Neaton_et_al:2005}. The vertical columns of black circles
represent various points of the polarization lattice for a fixed
amount of distortion, separated by the polarization quantum of
$185.6~\mu$C/cm$^2$. The dashed lines show how the polarization
lattice changes along the path.

The polarization lattice can be calculated as a gauge invariant geometric
phase of the wave-functions, the so-called ``Berry-phase''
\cite{King-Smith/Vanderbilt:1993,Vanderbilt/King-Smith:1993,Resta:1994}. This
expression solves the problem arising from the unboundedness of the quantum
mechanical position operator with the infinitely extended periodic
wave-functions in the solid and is compatible with the ``multivaluedness'' of
the polarization; the ambiguity of the polarization modulo a quantum
translates into an ambiguity of the Berry-phase modulo integer multiples of
$2\pi$.

In addition to ``beyond LSDA'' methods like pseudo-SIC and LSDA+$U$,
which result in stable insulating solutions for most magnetic oxides,
and the modern theory of polarization, which allows an accurate
calculation of experimentally measurable polarization differences,
there are many other developments in first-principles-based methods
that are relevant to the study of multiferroic materials, which we
mention briefly. For example, many multiferroic
systems have complex noncollinear spin structures, and coupling
effects between the polar and magnetic degrees of freedom can be due
to spin-orbit coupling. Therefore, a fully noncollinear treatment of
the spinor wave-functions and the inclusion of spin-orbit coupling in
the calculations is essential to describe many of the interesting
properties of multiferroic materials.  Also, the analysis of
ferroelectric instabilities using calculated phonon spectra is a very
powerful tool to explain the appearance of multiferroic properties and
predict the existence (or absence) of such instabilities in
designing new magnetoelectric multiferroic materials.

\section{First-principles explanation of experimental observations}

We now outline recent first-principles work that has led to
substantial progress in the understanding of {\it known} multiferroic
materials. First we describe studies aimed at understanding the
unusual mechanism for ferroelectricity in the hexagonal manganite
materials. Then we discuss a range of seemingly contradictory
experimental reports of both the magnetic and ferroelectric data in
multiferroic BiFeO$_3$, and show how first-principles calculations
were able to resolve some of the controversy.

\subsection{Elucidating the origin of ferroelectricity in hexagonal manganites}
\label{YMnO3}

Hexagonal YMnO$_3$ is the prototype of the class of isostructural
multiferroic materials with the chemical formula $R$MnO$_3$
($R$=Ho-Lu, Y) and room temperature space group $P6_3cm$. These
materials are ferroelectric with high ferroelectric Curie temperatures
of $\sim$ 1000~K, and exhibit noncollinear antiferromagnetic order
below $T_N$ $\sim$ 100~K. Although there have been many recent
experimental studies of the corresponding rare-earth systems,
primarily HoMnO$_3$, which show a complicated phase diagram at low
temperatures, the theoretical first-principles work has focused on
YMnO$_3$, partly because of the absence of 4$f$ electrons in this
system, which are difficult to treat using pseudopotential-based
methods. The main effort has been directed to understanding the
mechanism for the occurrence of ferroelectricity, which is clearly of
an unconventional origin, since the Mn$^{3+}$ cation does not have a
$d^0$ electron configuration and Y$^{3+}$ does not have a lone pair,
so the system contains no ``ferroelectrically active'' ions. Initial
progress in understanding the ferroelectric behavior was hampered by
the severe underestimation of the gap within LSDA
\cite{Filippetti/Hill:2002}, and by the fact that early experiments
incorrectly reported an off-centering of the Mn ions from the centers
of their polyhedra in the ferroelectric phase \cite{Yakel_et_al:1963}.

The first important contribution from first-principles calculations
was the analysis of the electronic structure of hexagonal YMnO$_3$
using the LMTO-ASA method \cite{Medvedeva_et_al:2000}. These
calculations, using the early experimentally reported paraelectric and
ferroelectric structures, showed that the Mn$^{4+}$ ion in hexagonal
YMnO$_3$ is not a Jahn-Teller ion despite of its $d^4$ electronic
configuration, because the trigonal crystal field splitting in this
material does not lead to an orbital degeneracy
\cite{Medvedeva_et_al:2000}. The width of the band gap was shown to be
very sensitive to the magnetic ordering, increasing (within the LSDA)
from about 0~eV for ferromagnetic \cite{Medvedeva_et_al:2000} and
A-type antiferromagnetic \cite{Qian/Dong/Zheng:2000} order to 0.47~eV
for a more complicated but still collinear antiferromagnetic spin
arrangement \cite{Medvedeva_et_al:2000}. The width of the gap was also
shown to strongly increase if correlations were incorporated in the
calculations through an LSDA+$U$ treatment
\cite{Qian/Dong/Zheng:2000,Medvedeva_et_al:2000}. This led to the
conclusion that both magnetic ordering and correlation effects have a
strong influence on the electronic structure, and play an important
role for the gap formation in hexagonal YMnO$_3$.  These calculations
also showed, however, that the overall electronic structure is {\it
not} strongly affected by the experimentally reported structural 
distortions leading from
the high temperature paraelectric phase to the low-temperature
ferroelectric phase. The only significant change observed in the
calculations was a slightly stronger $p$-$d$ hybridization due to the
off-center displacement of the Mn ion within its surrounding oxygen
cage that was reported in the early experiments
\cite{Yakel_et_al:1963}.

Subsequent plane-wave pseudopotential calculations confirmed that such
displacements of the Mn cation along the hexagonal axis indeed lead to
a certain degree of rehybridization between O 2$p$ and the Mn
3$d_{z^2}$ orbitals which are oriented along the polar $c$ direction,
whereas displacements within the $c$ plane are not accompanied by
large changes in chemical bonding
\cite{Filippetti/Hill:2001,Filippetti/Hill:2002}. These observations,
together with the experimental reports of Mn $c$-axis displacements,
led to the suggestion that a ``one-dimensional $d^0$-ness'', i.e. the
unoccupied nondegenerate $d_{z^2}$ orbital oriented along the
displacement direction, could be responsible for the stabilization of
the ferroelectric state \cite{Filippetti/Hill:2002}. Nevertheless, an
energy-lowering due to the displacement of the Mn ions was not found
in the first-principles calculations. It was unclear at this stage
whether the absence of energy lowering was an artifact of 
the use of the LSDA, which does not lead to an appropriate description
of the electronic structure of YMnO$_3$ (in fact, in the plane-wave
pseudopotential calculations the system remained metallic even in the
case of A-type antiferromagnetic ordering), or an indicator that the
early experiments reporting a Mn $c$-axis displacement were incorrect.
Therefore, a conclusive picture of the ferroelectricity could not be
achieved.

\begin{figure}
\centerline{
\includegraphics[width=0.25\textwidth]{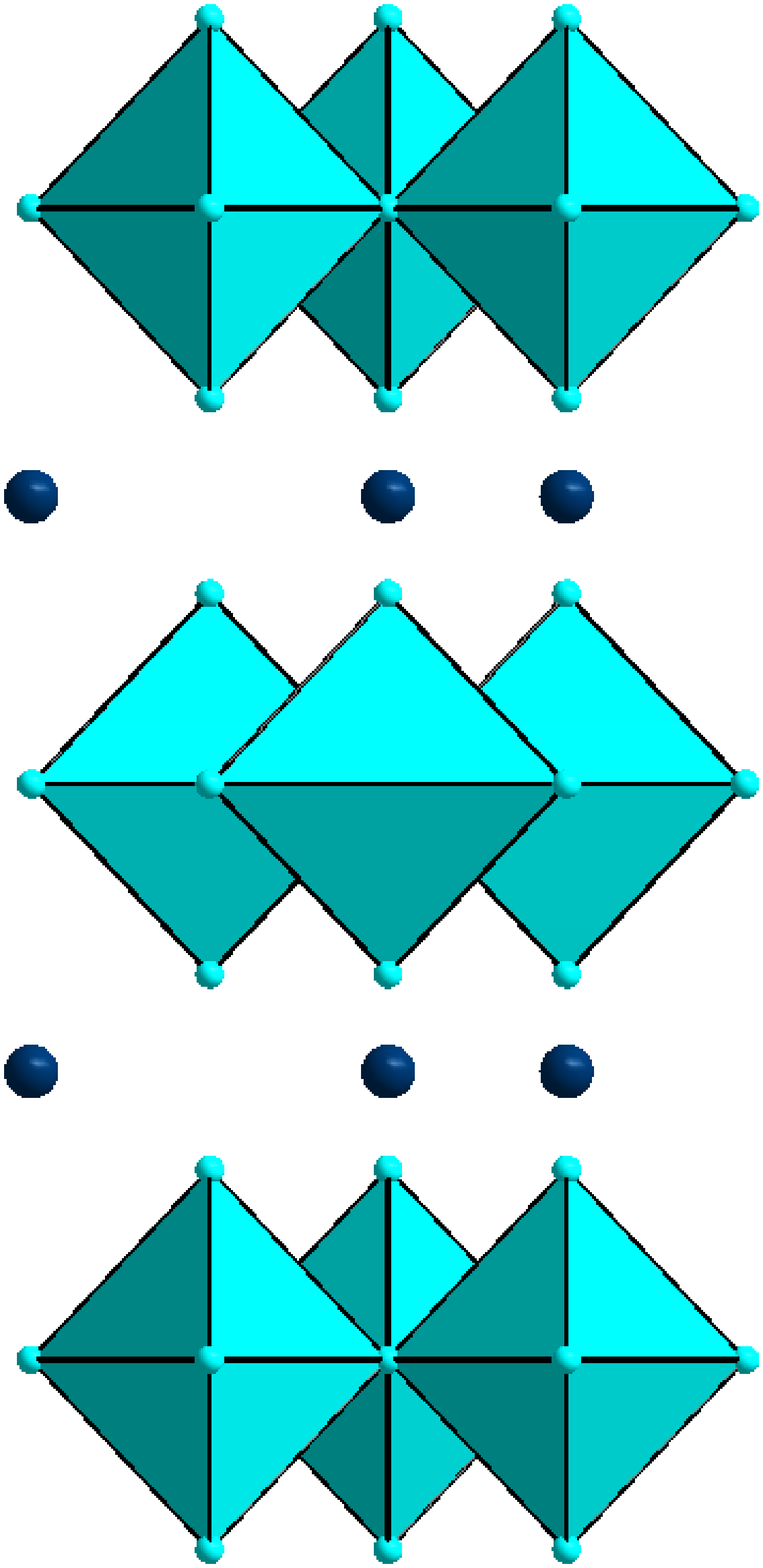} \hspace*{1cm}
\includegraphics[width=0.25\textwidth]{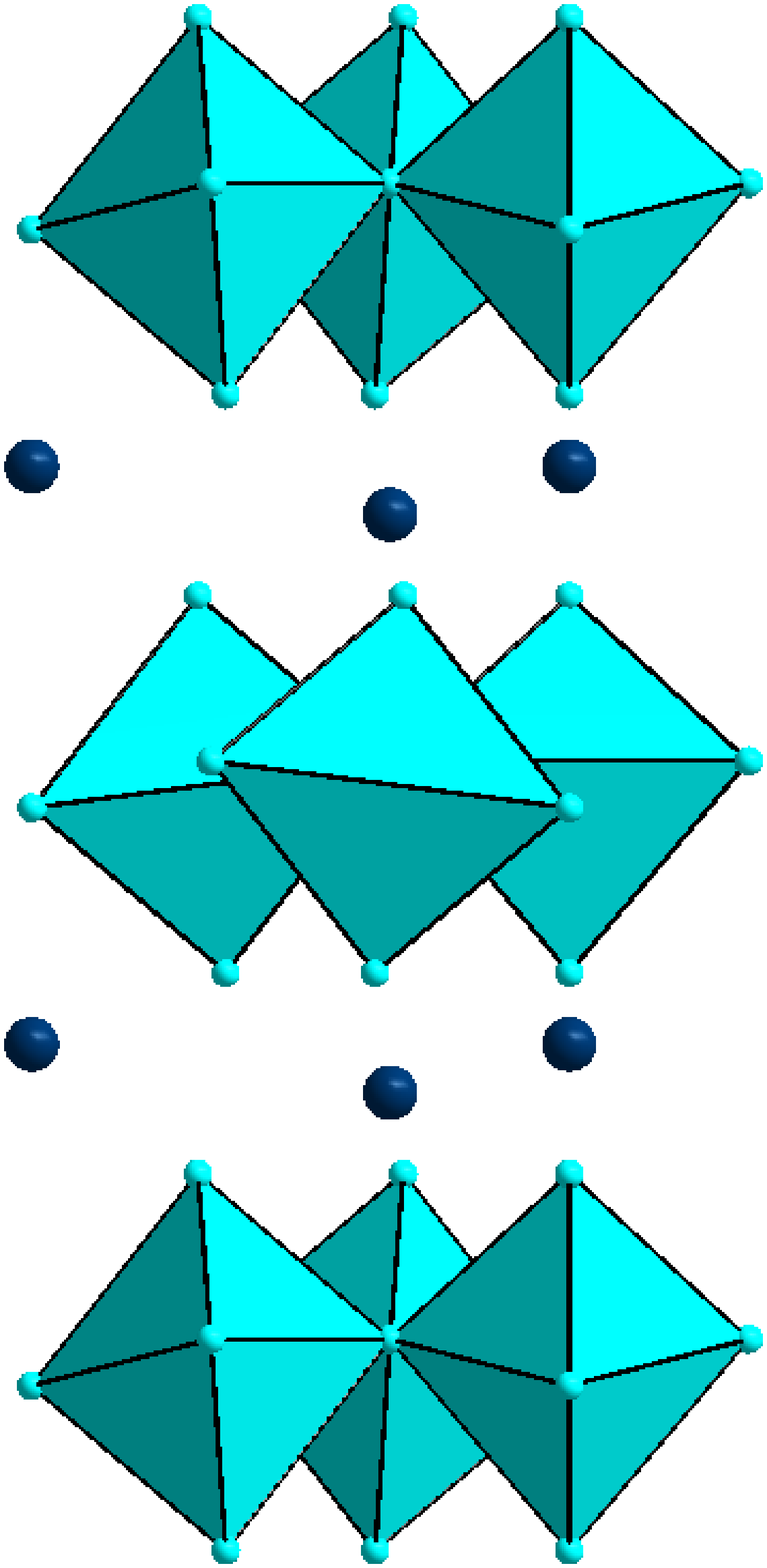}
}
\caption{High temperature centrosymmetric (left) and low-temperature
  ferroelectric structure (right) of YMnO$_3$. In the low-temperature
  structure the oxygen polyhedra (turquoise) undergo a collective
  rotation that leads to a unit cell tripling. This unit cell tripling
  is accompanied by displacements of the Y cations (blue) along the
  (0001) directions which lead to an electric dipole moment.}
\label{fig:YMnO3}
\end{figure}

The first breakthrough came through a collaboration that combined
progress on both the experimental and the theoretical side. First, new
single crystal x-ray diffraction experiments, using reflections of the
entire Ewald sphere, did not find an off-centering of the Mn ions
within their oxygen cages
\cite{VanAken/Meetsma/Palstra:2001}. Instead, the experimental data
indicated that the Mn ions remain in the center of their oxygen
polyhedra which instead undergo a collective rotation that leads to
unit cell tripling. In addition, the Y ions move along the hexagonal
$c$ axis, and fill the space that is freed due to the rotation of the
oxygen bipyramids (see Figure~\ref{fig:YMnO3}).  Second, on the
theoretical side the development of the self-interaction corrected
pseudopotential (pseudo-SIC) method \cite{Filippetti/Spaldin:2003},
motivated by the inadequacies of the LSDA, enabled a full structural
optimization of YMnO$_3$ while giving a realistic description of the
underlying electronic structure. These calculations confirmed the new
experimental data, giving structural parameters very close to those
obtained experimentally. The first-principles calculations also showed
that the structural distortions in the low temperature phase do not
lead to any significant charge redistribution. This is reflected in
the values of the Born effective charges, which were found to be very
close to the formal charges. It became clear that rehybridization is
not responsible for stabilizing the ferroelectric state in YMnO$_3$
and that in fact the interplay between polar and nonpolar structural
modes is essential to the ferroelectricity in this system
\cite{vanAken_et_al:2004}.

A concise picture of this interplay between polar and nonpolar
structural modes was obtained very recently by calculating and
analyzing the phonon instabilities in the high temperature
centrosymmetric structure \cite{Fennie/Rabe_YMO:2005}. It was shown
that the only unstable phonon mode is a unit-cell tripling mode, which
results in the experimentally observed collective tilting of the
oxygen polyhedra and a buckling of the Y (0001) planes ($K_3$
mode). Although this mode on its own does not lead to an electric
dipole moment, it reduces the symmetry of the system to the polar
space group $P6_3cm$, which corresponds to the experimentally observed
low-temperature structure. If the distortions that lead from the
high-temperature centrosymmetric to the low-temperature ferroelectric
phase are decomposed into contributions from different phonon modes,
the contribution of this $K_3$ mode makes up for more than 80~\% of
the final distortion. The second largest contribution (about 15~\%) is
due to a zone-center polar mode ($\Gamma_2^-$ mode) which is
compatible with the symmetry reduction due to the unstable $K_3$ mode
but is stable in the centrosymmetric high-temperature structure. By
mapping out the energy surface around the prototype phase as a
function of the $K_3$ and the polar $\Gamma_2^-$ mode it was shown
that the polar mode remains stable for all values of $K_3$ but that
due to a symmetry allowed coupling term between the $\Gamma_2^-$ and
$K_3$ modes the equilibrium value of the $\Gamma_2^-$ distortion
amplitude is shifted to a nonzero value. The $K_3$ mode therefore acts
as a ``geometric field'' on the $\Gamma_2^-$ mode. This behavior is
indicative of an \emph{improper ferroelectric} where the polar mode is
not the primary order parameter and this mechanism has been suggested
to be the driving force behind the ferroelectricity in YMnO$_3$
\cite{Fennie/Rabe_YMO:2005}.

The example of YMnO$_3$ shows how first-principles calculations of
phonon spectra and analysis of the calculated electronic structure can
give a good explanation of the mechanism driving the ferroelectric
phase transition in multiferroic materials. It is also apparent that
the use of the LSDA+$U$ and pseudo-SIC ``beyond LDA'' methods, and the
resulting realistic description of the electronic structure,
has been essential for such progress to be achieved.

\subsection{Resolving the true magnetization and polarization in BiFeO$_3$}

The determination of the spontaneous polarization and magnetization in
BiFeO$_3$ is an example of how valuable first-principles calculations
are for determining intrinsic material properties and explaining
contradictory experimental data. In a real sample, effects such as 
epitaxial strain, defects, microstructure, or interface properties 
can alter the physical properties of a material. Therefore it
is sometimes extremely difficult to determine the intrinsic material
properties unambiguously using only experimental data. First-principles
calculations on the other hand can deliver reference data
corresponding to a perfect sample containing absolutely no
defects. After establishing the intrinsic value of a certain quantity,
these calculations can also be used to determine how effects such as
epitaxial strain, various kinds of defects, or the presence of
interfaces and surfaces change this intrinsic value.

\begin{figure}
\centerline{\includegraphics[width=0.5\textwidth]{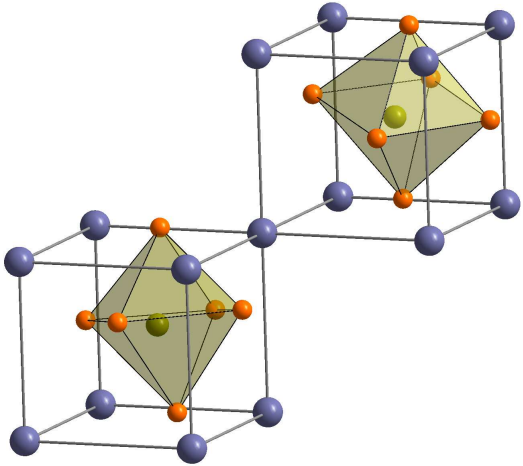}}
\caption{BiFeO$_3$ crystallizes in a distorted perovskite structure with space
  group $R3c$. The cations are displaced from their ideal positions relative to
  the anions along the [111] direction. In addition, the oxygen octahedra are
  rotated around the [111] direction, alternately clockwise and
  counterclockwise (Bi: blue, Fe: green, O: red).}
\label{fig:BiFeO3}
\end{figure}

Perovskite-structure BiFeO$_3$ is one rare example of a magnetoelectric
multiferroic that exhibits both magnetic and ferroelectric ordering above room
temperature. The bulk material becomes ferroelectric below $T_C \approx
1103$~K \cite{Teague/Gerson/James:1970}, adopting the $R3c$ structure shown in
Fig.~\ref{fig:BiFeO3}. The magnetic moments of the Fe cations order
antiferromagnetically (G-type) below $T_N \approx 643$~K
\cite{Kiselev/Ozerov/Zhdanov:1963,Fischer_et_al:1980}, and in addition it
exhibits a long-wavelength spiral-spin structure with a wavelength of about
620~\AA\ \cite{Sosnowska/Peterlin-Neumaier/Streichele:1982}. Until recently,
the value of the polarization in BiFeO$_3$ was believed to be rather
small. The only current-voltage measurements performed on BiFeO$_3$ single
crystals reported a spontaneous polarization of 3.5~$\mu$C/cm$^2$ along the
[100] direction (corresponding to 6.1~$\mu$C/cm$^2$ along the polar [111]
axis), but the corresponding hysteresis loops were far from being saturated
\cite{Teague/Gerson/James:1970}. Similar small values of the polarization were
subsequently reported for (Bi$_{0.7}$Ba$_{0.3}$)(Fe$_{0.7}$Ti$_{0.3}$)O$_3$
films on Nb-doped SrTiO$_3$ (100) \cite{Ueda/Tabata/Kawai:1999} and for
polycrystalline BiFeO$_3$ films \cite{Palkar/John/Pinto:2002}. However, all
these measurements were hampered by the high conductivity of the available
samples.

Interest in BiFeO$_3$ has grown considerably over the last few years, 
following a report of large electric polarization and substantial 
thickness-dependent magnetization in epitaxial thin films
\cite{Wang_et_al:2003}. The large ferroelectric polarization and macroscopic 
magnetization were unexpected, since all earlier studies had reported only 
a small value of the polarization and no magnetization. Therefore the large 
polarization measured in the epitaxial thin films of BiFeO$_3$ was initially 
ascribed to the effect of the heteroepitaxial constraint and the resulting 
change in the lattice parameters. Since a multiferroic with large polarization 
and magnetization above room temperature would be very desirable for 
technological applications \cite{Fiebig:2005}, a flurry of experimental 
activity ensued but could not resolve the confusion, with a range of reported
polarizations ranging between 2.2 $\mu$C/cm$^2$ \cite{Palkar/John/Pinto:2002} 
and 158 $\mu$C/cm$^2$ \cite{Yun_et_al:2004}.

The spread in magnetization and polarization data therefore posed a
number of questions:
\begin{enumerate}
\item What are the intrinsic values of the spontaneous polarization
  and the magnetization in BiFeO$_3$?
\item How are the polarization and the magnetization in BiFeO$_3$
affected by epitaxial strain?
\item How are the polarization and magnetization affected by defects?
\end{enumerate}
In the following we show how all these questions have been answered using
first-principles calculations.

\subsubsection{Intrinsic polarization of BiFeO$_3$}  

In Ref. \cite{Wang_et_al:2003} the large polarization found in the
epitaxial thin films of BiFeO$_3$ was initially ascribed to the effect
of the heteroepitaxial constraint and the resulting change in the
lattice parameters. To test this hypothesis, the polarization was
calculated from first-principles, both in the bulk rhombohedral
structure, and a hypothetical tetragonal structure, with its in-plane
lattice parameters constrained to that of the substrate. These first
calculations seemed to corroborate the hypothesis that the large
polarization was due to the epitaxial constraint, since the calculated
values for the two different structures were considerably
different. However, the LSDA was used in these calculations and no
insulating centrosymmetric reference structure (see
Section~\ref{sec:method}) could be obtained. Therefore, only the
\emph{absolute values} of the polarizations were reported (that is the
values at +100\% and $-$100\% distortion in Fig.~\ref{fig:path_endpt})
and not the \emph{differences} in polarization between the
ferroelectric structure and a centrosymmetric reference structure,
which would correspond to the polarization measured in a ferroelectric
switching experiment.  Note also, that subsequent experiments showed
that the structure in the films is in fact similar to the rhombohedral
bulk structure, albeit with an additional monoclinic distortion in the
case of a (001) oriented substrate
\cite{Li_et_al:2004,Qi_et_al:2005/2,Xu_et_al:2005}.

Subsequent first-principles calculations using the LSDA+$U$ method were able
to obtain an insulating solution for BiFeO$_3$ in the centrosymmetric cubic
perovskite structure and in the $R\bar{3}c$ structure, both of which are
reasonable reference structures for calculating the spontaneous polarization
of BiFeO$_3$ in the rhombohedral bulk structure
\cite{Neaton_et_al:2005}. Calculation of the polarization change along the
so-defined switching path resulted in a value for the spontaneous polarization
in bulk BiFeO$_3$ of about 95~$\mu$C/cm$^2$ along the [111] direction,
consistent with the values measured in the thin films
\cite{Wang_et_al:2003,Li_et_al:2004}, but exceeding the measured values in the
bulk samples by one order of magnitude. This result suggested that the small
values of the polarization reported in the past were in fact due to incomplete
switching and the resulting unsaturated hysteresis loops represented only a
fraction of the true polarization of BiFeO$_3$. The large polarization found
in the thin films is then the first really saturated measurement of the full
polarization of BiFeO$_3$.

\subsubsection{Effects of strain and defects on the spontaneous
  polarization in BiFeO$_3$}

It is known from materials such as BaTiO$_3$ and SrTiO$_3$ that the
epitaxial strain present in thin film samples can lead to drastic
changes in the ferroelectric polarization compared to the
corresponding bulk materials \cite{Choi_et_al:2004,Haeni_et_al:2004}.
The good agreement between the calculated bulk value of the
polarization and the experimentally observed thin film value in BiFeO$_3$ 
therefore seems to be in conflict with the strong strain dependence of
polarization expected in general for ferroelectrics.
The lattice mismatch between BiFeO$_3$ and SrTiO$_3$,
the substrate material used in Ref.~\cite{Wang_et_al:2003}, amounts to
about 2\% compressive strain in the BiFeO$_3$ film, which is a 
substantial value.

To address the epitaxial strain dependence of the ferroelectric polarization
in rhombohedral BiFeO$_3$ a detailed first-principles study was carried
out. Epitaxial strain was imposed by constraining the in-plane lattice
parameters over a range of non-equilibrium values, and allowing the structure
to relax out-of-plane at each constraint \cite{Ederer/Spaldin_2:2005}. In
contrast to similar calculations for BaTiO$_3$ and PbTiO$_3$
\cite{Neaton/Hsueh/Rabe:2002,Neaton/Rabe:2003,Bungaro/Rabe:2004}, it was found
that the electric polarization in BiFeO$_3$ is not significantly affected by
the presence of epitaxial strain. Even strain values as large as $\pm$ 3\%
lead only to changes in the polarization of a few percent, provided that the
material does not undergo a structural phase transition. (For comparison, the
corresponding change in BaTiO$_3$ is more than 100\%
\cite{Neaton/Hsueh/Rabe:2002}). The relative strain-independence of the
polarization in BiFeO$_3$ supports the notion that the large polarization
found in the thin films is essentially equal to the bulk value and that the
old measurements severely underestimated the polarization in BiFeO$_3$.

Recently, possible reasons for this different strain behavior between
BiFeO$_3$ and ferroelectrics like BaTiO$_3$ or PbTiO$_3$ have been
addressed using first-principles methods
\cite{Ederer/Spaldin_3:2005}. It was shown that the variation of the
spontaneous polarization with epitaxial strain in thin film
ferroelectrics can be understood in terms of the piezoelectric and
elastic constants of the corresponding bulk materials, and that a
large strain dependence is not a common feature of all ferroelectrics.

Ref.~\cite{Ederer/Spaldin_2:2005} also investigated the effect of
oxygen vacancies on the spontaneous polarization in BiFeO$_3$ using
first-principles techniques. It was shown that the incorporation of
such vacancies leads to small relaxations around the vacancy site and
to the formation of Fe$^{2+}$ next to the vacancy site, but that the
value of the spontaneous polarization is not significantly affected by
the presence of oxygen vacancies.

\subsubsection{Magnetic properties of BiFeO$_3$}

First-principles calculations have also been applied to understanding the
magnetic properties of BiFeO$_3$, in particular the observation of
magnetization in BiFeO$_3$ thin films and its absence in bulk samples. One
possible source of magnetization in a primarily antiferromagnetic material is
{\it weak ferromagnetism}, i.e.\ a canting of the mainly antiferromagnetically
oriented magnetic moments resulting in a small magnetization
\cite{Ederer/Spaldin:2005}. In bulk BiFeO$_3$, the G-type antiferromagnetic
ordering \cite{Fischer_et_al:1980} is modulated by a long wavelength spiral
spin structure \cite{Sosnowska/Peterlin-Neumaier/Streichele:1982} which would
lead to a cancellation of the magnetization on a macroscopic scale. However,
it was suggested that the thin film morphology might suppress the spiral spin
structure so that the weak ferromagnetism becomes observable in thin films
\cite{Ederer/Spaldin:2005,Bai_et_al:2005}. First-principles calculations
indeed showed that a local canting does occur, and that its origin is the
Dzyaloshinskii-Moriya interaction \cite{Moriya:1963}. The magnitude of the
calculated value for the magnetization is approximately
0.1~$\mu_\text{B}$/(unit cell), in good agreement with recent measurements
\cite{Bai_et_al:2005,Eerenstein_et_al:2005}. Therefore, this supports the
interpretation proposed above for the origin of the magnetization in thin
films of BiFeO$_3$.

The thin film measurements of Ref. \cite{Wang_et_al:2003} also reported a
strong increase in magnetization with decreasing film thickness, reaching
values of about 1~$\mu_\text{B}$/Fe for film thicknesses below 100~nm. Since
neither calculations on ideal systems \cite{Ederer/Spaldin:2005} nor recent
measurements on highly stoichiometric BiFeO$_3$ films
\cite{Eerenstein_et_al:2005} were able to reproduce an increased
magnetization, it was suggested that oxygen vacancies and the resulting
presence of Fe$^{2+}$ could lead to the observed increase in the magnetization
of very thin films \cite{Wang_et_al_2:2005}. However, supercell calculations
corresponding to different concentrations of oxygen vacancies were not able to
confirm this suggestion \cite{Ederer/Spaldin_2:2005}. In these calculations it
was found that removing a neutral oxygen atom from the supercell resulted in
the formation of Fe$^{2+}$ on the sites adjacent to the vacancy site. The
magnetic moments of such a pair of Fe$^{2+}$ cations are coupled
antiferromagnetically and therefore do not result in a net
magnetization. Also, the canting of the magnetic moments was not significantly
affected by the presence of the oxygen vacancies, so that no increase in the
macroscopic magnetization due to oxygen vacancies could be
found. Ref.~\cite{Ederer/Spaldin_2:2005} also showed that the canting, and
with this the magnetization, is not significantly enhanced by epitaxial
strain. The reason for the increase in magnetization with decreasing film
thickness in Ref.~\cite{Wang_et_al:2003} is therefore still unclear, but from
the first-principles calculations it can be concluded that the observed effect
must be caused by effects other than epitaxial strain or neutral oxygen
vacancies.

\subsection{Other systems}

Very recently, phonon energies were calculated for the spinel system
CdCr$_2$S$_4$ \cite{Fennie/Rabe_CCS:2005}. This material, which is known to be
a rare example of a ferromagnetic insulator with a rather high magnetic Curie
temperature, was recently reported to exhibit interesting coupling between
dielectric and magnetic properties as well as a small electric polarization at
low temperatures \cite{Hemberger_et_al:2005}. It was suggested that the
relaxor-like behavior could be due to frustrated soft-mode behavior and that
the ferroelectricity at low-temperatures could originate from a displacement
of the Cr$^{3+}$ ion \cite{Hemberger_et_al:2005}. The first-principles
calculation on the other hand showed that all phonon modes in this material
are rather stable and that the scenario of a Cr$^{3+}$ ion moving off-center
is extremely unlikely \cite{Fennie/Rabe_CCS:2005}. More recent experimental
results indicate that the observed behavior is probably defect-related and not
an intrinsic effect since the observed features in the dielectric constant
vanish after annealing the corresponding samples
\cite{Hemberger_et_al_2:2005}. This represents another example how
first-principles calculations can be used to separate intrinsic material
properties from extrinsic effects like microstructure or defects.

\subsection{Summary of first-principles explanation of experimental
  observations}

It is clear from the above examples that first-principles calculations
have made many contributions to our understanding of multiferroic
phenomena, and that the interplay between theory and experiment has
been invaluable for progress in the field.  The example of YMnO$_3$
showed how the mechanism underlying unconventional ferroelectric phase
transitions can be elucidated using first-principles techniques, and
the example of BiFeO$_3$ showed how first-principles calculations can
be used for resolving conflicting experimental results and to separate
intrinsic materials properties from extrinsic effects such as defects or
microstructure.

\section{First-principles predictions of new materials and novel phenomena}

In this section we change our focus compared to the previous section
and discuss examples where predictions have been made using
first-principles techniques prior to their experimental verification.

\subsection{Electric-field induced magnetization switching}

The most important question arising from the presence of both
spontaneous electric polarization and macroscopic magnetization in a
multiferroic is whether these two properties are coupled, and, if so, in 
what manner. A coupling that facilitates the reversal of the magnetization 
by an electric field or conversely, the reversal of the electric polarization 
by a magnetic field, would be of particular interest. In the
following we show how first-principles calculations have been used to
study this question, and to develop theoretical models for coupling
between magnetic and structural/dielectric properties.

Again, BiFeO$_3$ has been used as the prototype material for 
first-principles study. Calculations of the coupling between magnetism
and structural distortions in BiFeO$_3$ showed that the
Dzyaloshinskii-Moriya interaction \cite{Moriya:1963}, which leads to
the canting of the magnetic moments and the appearance of a
macroscopic magnetization, is indeed a result of the symmetry lowering
structural distortions in this material (in the high symmetry cubic
phase, weak ferromagnetism is not allowed by symmetry)
\cite{Ederer/Spaldin:2005}. In the case of BiFeO$_3$ there are two
different symmetry lowering distortions that together reduce the
symmetry from the cubic perovskite structure ($Pm\bar{3}m$) to the
ferroelectric rhombohedral ground state structure ($R3c$). One is a
nonpolar $R$ mode that leads to a unit cell doubling compared to the
simple perovskite structure and rotations of the oxygen octahedra
around the rhombohedral axis, and the other is a polar $\Gamma$ mode
that displaces all ions relative to each other along the same
rhombohedral axis (see Figure~\ref{fig:BiFeO3}). It was shown that the
Dzyaloshinskii-Moriya interaction in BiFeO$_3$ is caused by the
nonpolar mode and that reversal of that mode also inverts the
magnetization direction \cite{Ederer/Spaldin:2005}.

\begin{figure}
\centerline{\includegraphics[width=0.7\textwidth]{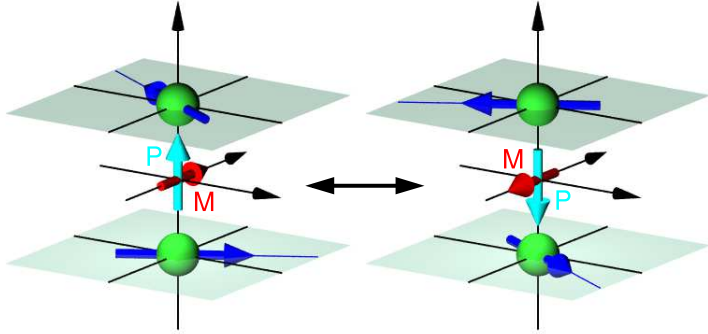}}
\caption{Possible realization of electric-field induced magnetization
  switching in a weak ferromagnetic ferroelectric: if the polarization
  $P$ is reversed, the canting of the antiferromagnetic sublattices,
  and with this the magnetization $M$, is also reversed. }
\label{fig:switching}
\end{figure}

This has important consequences for the possible realization of
electric-field-induced magnetization switching, which was believed to
be impossible due to symmetry reasons \cite{Schmid:1999}. From the
results obtained in Ref.~\cite{Ederer/Spaldin:2005} it became clear
that such an effect is in principle possible in weak ferromagnetic
systems in which the polar distortion, which results in
ferroelectricity, also leads to the appearance of a
Dzyaloshinskii-Moriya interaction, which results in weak
ferromagnetism. In such a scenario, the magnetization will be strongly
coupled to the ferroelectric polarization and reversal of the
ferroelectric polarization by an electric field will be accompanied by
reversal of the macroscopic magnetization \cite{Ederer/Spaldin:2005}
(see Figure~\ref{fig:switching}).  The search for a material where this
effect can be realized is a topic of current research activity
\cite{Ederer/Spaldin:unpublished}.

\subsection{Computational materials design}

The examples discussed in the previous sections illustrate that
first-principles calculations are capable of correctly predicting
numerous properties of magnetoelectric multiferroics. This opens up
the possibility of \emph{ab initio} computational design of entirely
new multiferroic materials, with improved properties, in advance of
their synthesis. Such a materials design approach was suggested in
Ref.~\cite{Spaldin/Pickett:2003} and has been applied since then to
search for new multiferroics with properties superior to those of
existing materials.

The scarcity of magnetoelectric multiferroics has rendered such an
approach extremely desirable. For example, most magnetoelectric
multiferroics are either antiferromagnets or exhibit only weak
ferromagnetism, often below room temperature. On the other hand, for
technological applications a multiferroic with large polarization and
large magnetization, well above room temperature, and preferably with
strong coupling between the two properties is needed. Currently, no
magnetoelectric material is known that fulfills all these
requirements. Since the calculation of the properties of a variety of
candidate materials is often less costly than the corresponding
experimental synthesis and subsequent characterization, a
first-principles design approach is very appealing.

Probably the first example for the {\it ab initio} design of a
magnetoelectric multiferroic was the prediction of ferroelectricity in
BiMnO$_3$. This material was known to be ferromagnetic at low
temperatures \cite{Bokov_et_al:1966}, and the calculation of unstable
zone-center phonon modes in cubic BiMnO$_3$ suggested the simultaneous
existence of ferroelectricity \cite{Hill/Rabe:1999}. Motivated by this
prediction, experimental investigations were performed and
ferroelectric hysteresis loops were reported for polycrystalline
samples as well as for thin films of BiMnO$_3$
\cite{dosSantos_et_al_SSC:2002}. The ferromagnetism in BiMnO$_3$ is
interesting in its own right, and is believed to be due to orbital
ordering caused by a cooperative Jahn-Teller distortion
\cite{dosSantos_et_al_PRB:2002}. Such an orbital ordering was
reproduced recently by first-principles calculations using the
experimentally obtained structure \cite{Shishidou_et_al:2004}.

A first-principles search for multiferroism has also been performed
for the closely related compound BiCrO$_3$
\cite{Hill/Baettig/Daul:2002}. Prior to the theoretical study, little
was known experimentally about the material; early papers reported
antiferromagnetism with a weak ferromagnetic moment below 123K, and a
structural transition between pseudomonoclinic and pseudotriclinic
perovskite structures at 400K \cite{Sugawara_et_al:1968}. The
first-principles calculations found an antiferromagnetic ground state,
in agreement with the Goodenough/Kanamori/Anderson rules for
superexchange interactions in magnetic insulators
\cite{Goodenough:Book,Anderson:1963}, and the calculation of the full
phonon spectrum revealed several unstable phonon branches over the
whole range of the Brillouin zone. The most unstable modes were found
at the $R$ and $M$ points of the Brillouin zone corresponding to
antiferroelectric or antiferrodistortive phonon modes. Since the
antiferromagnetic ordering in BiCrO$_3$ does not lead to a macroscopic
magnetization, it was concluded that this material will not be of
great technological relevance and a full structural optimization of
the ground state structure was not attempted
\cite{Hill/Baettig/Daul:2002}. A subsequent experimental study
\cite{Niitaka_et_al:2004} confirmed the weak ferromagnetism, and
observed a non-polar to polar structural phase transition, suggestive
of ferroelectricity, at 440K.

The difficulty of incorporating a substantial magnetization in a
magnetoelectric multiferroic results from the dominance of magnetic
superexchange in such transition metal oxides, which in most cases
leads to antiferromagnetic nearest neighbor coupling. Although in
certain cases the superexchange mechanism can lead to ferromagnetic
coupling, the strength of such a ferromagnetic coupling is usually
substantially weaker than in the antiferromagnetic case
\cite{Anderson:1963}, resulting in low magnetic ordering
temperatures. Therefore a recent proposal to exploit the strong
antiferromagnetic superexchange and introduce macroscopic
magnetization by combining two magnetic ions with different magnetic
moments, resulting in a \emph{ferrimagnetic} spin-arrangement
\cite{Baettig/Spaldin:2005} is very promising. The ordered double
perovskite system Bi$_2$FeCrO$_6$ was suggested as a candidate
material, since the stereochemically active Bi$^{3+}$ ion should
introduce a potentially ferroelectric structural distortion and the
closed-subshell $d^5$ Fe$^{3+}$ and $d^3$ Cr$^{3+}$ ions should result
in insulating behavior.

Indeed, a ferroelectric ground state with space group $R3$ was found by
structural optimization using first-principles calculations within the
LSDA+$U$ method \cite{Baettig/Spaldin:2005}. The corresponding structure is
closely related to the $R3c$ structure found in the well-known multiferroic
BiFeO$_3$ \cite{Neaton_et_al:2005} and is also consistent with the strong
phonon instability at the $R$ point found in cubic BiCrO$_3$
\cite{Hill/Baettig/Daul:2002}. A rock-salt-like ordering of Fe$^{3+}$ and
Cr$^{3+}$ cations on the $B$ site lattice of the perovskite structure was
enforced in the calculation; such ordered structures can be achieved by
layer-by-layer growth in epitaxial films, as has been demonstrated
experimentally for the case of La$_{2}$FeCrO$_{6}$
\cite{Ueda/Tabata/Kawai:1998}. The calculated spontaneous electric
polarization of Bi$_2$FeCrO$_6$ was about 80~$\mu$C/cm$^2$. The calculated
ground state magnetic structure was the ferrimagnetic equivalent of G-type
antiferromagnetic ordering, with oppositely oriented magnetic moments of the
Fe$^{3+}$ and Cr$^{3+}$ ions resulting in a magnetization of
2~$\mu_\text{B}$/formula unit. Thus, ordered Bi$_2$FeCrO$_6$ was predicted to
be a unique magnetoelectric multiferroic with large electric polarization and
large magnetization \cite{Baettig/Spaldin:2005}. The magnetic ordering
temperature of the newly predicted magnetoelectric multiferroic
Bi$_2$FeCrO$_6$, estimated by calculating the nearest and next-nearest
neighbor coupling constants from first-principles and then applying the
mean-field approximation, was shown to not exceed 100~K
\cite{Baettig/Ederer/Spaldin:2005}.  Since this mean-field value can be viewed
as an upper bound for the real magnetic ordering temperature it is unlikely
that Bi$_2$FeCrO$_6$ will show multiferroic behavior at room temperature.

In order to gain more insight into the factors that promote high magnetic
ordering temperatures in such ferromagnetic ferroelectrics, a comparative
study of the series of multiferroics BiFeO$_3$, Bi$_2$FeCrO$_6$, and BiCrO$_3$
was performed \cite{Baettig/Ederer/Spaldin:2005}. The magnetic ordering
temperatures of the different compounds were again obtained by calculating the
nearest and next nearest neighbor exchange coupling constants using first
principles methods and determining the resulting ordering temperature within
the mean-field approximation. It was found that the calculated variation of
the strength of the nearest neighbor coupling constants could be explained
based on the different electron configurations of the constituent ions using
Anderson's theory of superexchange \cite{Anderson:1963}. In addition, the
influence of the structural distortions on the magnetic coupling was
determined by comparing the calculated coupling constants for the relaxed
ferroelectric structures with those obtained for the corresponding undistorted
cubic structures. In general, the structural distortions resulted in a
weakening of the antiferromagnetic coupling, and in the case of
Bi$_2$FeCrO$_6$ an interesting crossover behavior from antiferromagnetic to
ferromagnetic nearest neighbor coupling could be observed, again in good
agreement with the qualitative predictions from the theory of superexchange
\cite{Baettig/Ederer/Spaldin:2005}.  The insight gained by studying the
systematic variations of exchange interactions in this series of compounds was
used to propose a variety of other systems with potentially larger magnetic
coupling. Since the theory of superexchange is only able to give qualitative
predictions, however, additional first-principles calculations will be
necessary to make quantitative predictions.

\section{Summary and Outlook}

From the examples discussed in the present paper it is clear that first
principles methods are very powerful tools for studying multiferroic materials
and magnetoelectric coupling. These methods facilitate the accurate
quantitative calculation of many material properties and can therefore be used
to establish benchmark values to verify experimentally obtained
data. Furthermore, the detailed analysis of the calculated results and the
possibility to perfectly control the system under investigation allows the
development of new theoretical concepts and the construction of
phenomenological models to gain a deeper understanding of the basic physics
behind multiferroic phenomena. This knowledge can then be used to design new
materials with desirable properties and specific values can be calculated
explicitly for various candidate materials. The examples discussed in the
previous sections illustrate how first-principles calculations have been used
to deliver benchmark data for the magnetization and polarization in BiFeO$_3$,
to understand the mechanism underlying the ferroelectricity in YMnO$_3$, to
establish the concept of electric-field induced magnetization switching in
weak ferromagnets, and to design new multiferroic materials with formerly
unachievable properties.

In the future, first-principles methods will undoubtedly continue to make
invaluable contributions to the field of magnetoelectric multiferroics.  For
example, efforts are underway to describe effects of spin-phonon coupling
\cite{Fennie/Rabe:unpublished}, and to identify a material that exhibits
electric-field switchable weak ferromagnetism
\cite{Ederer/Spaldin:unpublished}.  In addition, new classes of multiferroic
materials remain to be explored, for example the orthorhombic and hexagonal
rare earth manganites, which show a variety of interesting coupling effects
between magnetic and ferroelectric properties
\cite{Kimura_et_al_Nature:2003,Lottermoser_et_al:2004}. Nanostructured
magnetoelectric composites \cite{Zheng_et_al:2004} offer a new way of coupling
magnetic and ferro- or piezoelectric properties, and the multitude of possible
materials combinations represents a way to tune the properties of the
heterostructure to the desired values. Again, first-principles calculations
offer a powerful tool to study the properties of such heterostructures and to
predict the characteristics of different materials combinations. We therefore
believe that the first-principles study of multiferroic materials will
continue to be very exciting and that these methods will continue to
contribute significantly to the progress in this field.



\bibliographystyle{elsart-num} \bibliography{references.bib}







\end{document}